\documentclass[aps,prl,10pt,twocolumn,floatfix,groupedaddres]{revtex4-2}
\usepackage{graphicx}
\usepackage{epsfig}       
\usepackage{longtable}
\usepackage{dcolumn}
\usepackage{amsfonts}     
\usepackage{amsmath}      
\usepackage{amssymb}      
\usepackage{bm}           
\usepackage{color}
\usepackage[normalem]{ulem}
\usepackage[colorlinks=true, urlcolor=blue, linkcolor=blue, citecolor=blue]{hyperref}
\usepackage[all]{hypcap}

\newcommand{\beq}{\begin{equation}}
\newcommand{\eeq}{\end{equation}}
\newcommand{\bea}{\begin{eqnarray}}
\newcommand{\eea}{\end{eqnarray}}

\newcommand{\bk}{\mathrm{\textbf{k}}}

\newcommand{\bq}{\mathrm{\textbf{q}}}

\begin{document}

\title{\textbf{The magnon spectra of g-type altermagnet bulk CrSb}}

\author{Murat Tas}\email{murat.tas@gtu.edu.tr}

\affiliation{Department of Physics, Gebze Technical University, 41400 Kocaeli, Turkey 
}
\date{\today}

\begin{abstract}
We present a calculation of the magnon spectra and chiral lifetimes of altermagnons 
in bulk CrSb using the many-body perturbation theory. The spin-split band structure 
is evident in the magnon spectra. Altermagnons attain an energy of 275~meV at the K 
point of the Brillouin zone. Due to large spin splitting at a specific $\bq$ point 
along the A - M direction, lifetime differences of chiral altermagnons reach 
approximately 20~fs.
\end{abstract}
\maketitle

\section{Introduction} \label{intro}
Altermagnetism has been a newly recognized form of collinear magnetism since its 
first theoretical prediction \cite{Hayami2019,Yuan2020,Smejkal2020,Mazin2021,
Smejkal2022,Smejkal2022-2,Bai2024} followed shortly after by its experimental 
confirmations \cite{exp-RuO2,CrSb-film}. Altermagnets (AMs) possess an 
antiferromagnetic ground state, but the opposite-spin electrons are connected by 
rotational symmetry instead of translational or inversion symmetry as in 
conventional antiferromagnets (AFMs). Unlike AFMs, the AMs break time-reversal 
symmetry, or Kramers degeneracy, due to a remarkable spin-splitting observed in 
their nonrelativistic electronic band structure, a hallmark of ferromagnets (FMs), 
in certain parts of their Brillouin zone (BZ). Hence, they exhibit properties 
characteristic of both ferro- and antiferro-magnets, such as the anomalous Hall and 
magneto-optic effects, making them promising candidates for spintronics applications. 
Furthermore, there has been active research to realize the superconducting diode 
effect in altermagnets \cite{Banerjee-SDE}, and to investigate the interplay between 
altermagnetism and p-wave superconductivity \cite{chatterjee2025}. Hence, there has 
been immense theoretical and experimental research for identifying AMs \cite{2d-AMs}. 
Currently, there are only a few confirmed AMs such as RuO$_2$, MnTe, CrSb and 
Mn$_5$Si$_3$ \cite{Reichlova2024}.

To comprehend dynamical properties of a magnetic system, it is essential to 
investigate its elementary spin excitations, which range from low-energy collective 
spin-wave excitations, or magnons, to high-energy single-particle Stoner excitations. 
Hence, spin excitations are present across all temperatures and significantly 
influence the physical properties of magnetic materials, such as specific heat and 
magnetic susceptibility. Understanding the nature of magnons in AMs, called 
altermagnons, is crucial for development of ultrafast magnetic storage and 
processing devices. Consequently, this subject has become an active research field 
\cite{magnons-in-RuO2,Costa2024}.

The magnon dispersion near the $\Gamma$ point exhibits a remarkable diversity across 
magnetic materials. FMs are characterized by a quadratic dispersion, while AFMs 
display a linear dependence. AMs, however, possess a unique spin structure that 
gives rise to chiral altermagnons.

In this article, we report the spectra and lifetimes of altermagnons in bulk CrSb 
computed using the many-body perturbation theory (MBPT), which is applicable to 
systems with both localized and itinerant magnetic moments.

Both bulk and film forms of CrSb have been extensively investigated \cite{Snow1952,
Park2020,CrSb-film,Meng,Ding}. Bulk CrSb crystallizes in the hexagonal (hcp) 
NiAs-type structure with space group P$6_3$/mmc (No. 194). Its primitive unit 
contains two Cr and two Sb atoms. The Cr atoms have different environment and are 
interconnected by six-fold screw rotation symmetry. While the nearest-neighbor Cr 
atoms are aligned antiferromagnetically, the second-nearest ones are aligned 
ferromagnetically. Bulk CrSb exhibits a N\'eel temperature of approximately 705~K, 
implying a stable antiferromagnetic order at relatively high temperatures. 
Li \emph{et al}. \cite{Li2024} observed a large momentum-dependent spin-splitting 
of approximately 1~eV in bulk CrSb via high resolution angle-resolved photoemission 
spectroscopy (ARPES) and spin-ARPES measurements. They concluded CrSb as a Weyl 
semimetal hosting two distinct types of spin-carrying nodal structures. Employing 
the spin-integrated soft X-ray ARPES, Reimers \emph{et al}. \cite{CrSb-film} 
measured the band dispersions of epitaxial thin films and confirmed the band 
structure calculations that predict CrSb as an AM. Ding \emph{et al}. have recently 
discovered spin splitting of 0.93~eV near the Fermi level and a distinctive g-wave 
spin texture in momentum space in bulk CrSb using ARPES \cite{Ding}.

We obtained the equilibrium lattice parameters for bulk CrSb as $a=b=4.108$~\AA~and 
$c=5.440$~\AA, demonstrating excellent agreement with the experimental values of 
$a=4.127$~\AA~and $c=5.451$~\AA~reported by Snow \cite{Snow1952}. The ground state 
and nonrelativistic electronic band structure are computed with FLEUR code 
\cite{Fleurcode}, which implements the all-electron full-potential linearized 
augmented plane-wave (FLAPW) method within the DFT. The Perdew-Burke-Ernzerhof (PBE) 
\cite{PBE} parametrization of the generalized-gradient approximation is used to 
calculate the exchange-correlation energy. For the valence electrons, we set an 
angular momentum cutoff of $l_\mathrm{max}=10$ within the muffin-tin (MT) spheres 
and a plane-wave cutoff of 4.5~bohr$^{-1}$ in the interstitial region. The MT radii 
for Cr and Sb atoms are, respectively, 2.51 and 2.53~bohr. The BZ is sampled using 
a $\bk$-point grid of $24 \times 24 \times 18$.

\begin{figure}[!hbt]
\includegraphics[scale=0.7]{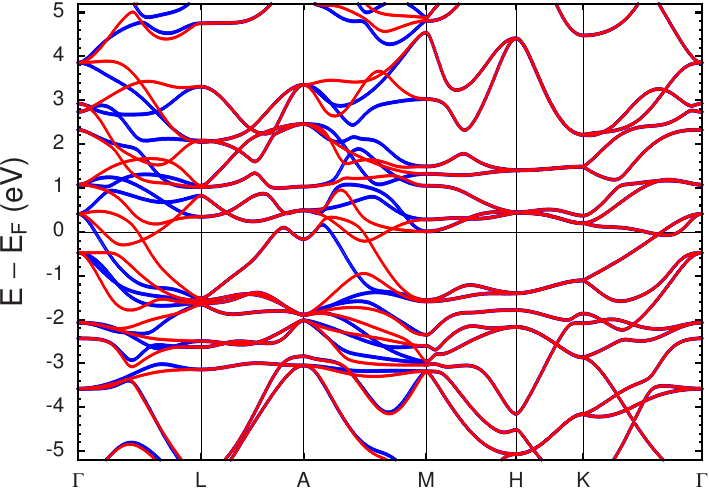}
\caption{The nonrelativistic electronic band structure of bulk CrSb. We observe 
a remarkable energy difference between majority (red) and minority (blue) spin 
bands along the $\Gamma$ - L and A - M directions.}
\label{band-CrSb}
\end{figure}

The ground state of bulk CrSb is found to be an antiferromagnetic metallic phase 
with magnetic moments of $\pm\,2.79$~$\mu_\mathrm{B}$ localized on the Cr atoms. Its 
electronic band structure presented in Fig.~\ref{band-CrSb} exhibits a significant 
energy difference between electrons with majority and minority spins along the 
$\Gamma$ - L and A - M directions. These spin-split bands provide a strong evidence 
that bulk CrSb exhibits an altermagnetic behavior.

\section{The magnon spectra and spin lifetime} \label{magnon}
The physics of spin waves evidently comprises an unusually rich area of research. 
A wealth of information regarding the spin dynamics in solids can be gleaned from 
the transverse magnetic response function, or transverse magnetic susceptibility 
(TMS), which is defined as \cite{Friedrich2019}
\bea
\chi^{-+}(\bq,\omega)=\frac{\chi_\mathrm{KS}^{-+}(\bq,\omega)}
{1 - W^{-+}(\bq,\omega)\chi_\mathrm{KS}^{-+}(\bq,\omega)}\,.
\eea
$\chi^{-+}$ ($\chi^{+-}$) describes component of the response of a magnetic material 
that is perpendicular to an external magnetic field for spin-flip transitions from 
down to up (up to down) state. $\chi^{-+}_\mathrm{KS}$ represents the noninteracting 
(Kohn-Sham) magnetic response, and $W$ is the fully screened Coulomb potential. The 
short-range nature of $W$ in metallic systems allows us to treat the electron-hole 
interactions using an on-site approximation only on the same atom and static limit 
of $W$.

The imaginary part of $\chi^{-+}_\mathrm{KS}$ directly measures the single-particle 
spin-flip Stoner excitations, whereas that of the denominator provides magnon 
dispersion. Consequently, the TMS serves as a central quantity for theoretical 
description of the magnetic order.

Paischer \emph{et al}. \cite{Paischer} recently investigated impact of the 
electron-magnon interaction on the electronic properties of bulk CrSb. Their 
findings suggest that strength of the electron-magnon interaction is significantly 
influenced by the density of available final particle states in the scattering 
processes, as well as by the spatial characteristics of coupling magnonic modes. 
Recently, Biniskos \emph{et al}. \cite{Biniskos2025} demonstrated a direct 
connection between the circular dichroism of the magnon peak detected in resonant 
inelastic X-ray scattering (RIXS) and predicted energy splitting of the two 
circularly-polarized magnons in CrSb. Their results confirm the theoretical 
predictions and the g-wave symmetry in CrSb.

A fat-band analysis of the bulk CrSb band structure indicates that bands near the 
Fermi level originate from the hybridization of Cr $3d$ and Sb $4p$ orbitals. We 
thus utilize these, in total of 16, orbitals as the projectors when expanding $W$ 
in terms of products of MLWFs. This approach enables us to compute the TMS at 
arbitrary $\bk$ points. The MLWFs are constructed using WANNIER90 code 
\cite{Wannier90}, which is interfaced with SPEX code.

To ensure high-precision calculation of $W$, we employed a dense $\bk$-point mesh 
of $12 \times 12 \times 10$, comprising 160 $\bk$ points in the irreducible BZ, and 
included 500 bands in the calculation. $W$ is expanded in a mixed product basis set 
that incorporates contributions from both the MT spheres and plane waves in the 
interstitial region. For this mixed product basis set, we used the cutoff parameters 
of $L_\mathrm{max}=5$ and $G_\mathrm{max}=4$~bohr$^{-1}$. Interaction matrix 
elements were then employed for the computation of magnon dispersions. The 
Brillouin-zone summations in $\chi^{-+}_\mathrm{KS}$ were performed with the 
tetrahedron method.

We computed the spectra of altermagnons in bulk CrSb within the framework of MBPT, 
as implemented in SPEX code \cite{Spex,Spexcode}. The implementation incorporates 
coupling of electrons to holes with opposite spins through $W$, which is calculated 
within the constrained random phase approximation (cRPA) using the Kohn–Sham 
electronic bands and wave functions. This interaction is accounted for by vertex 
corrections in the form of ladder diagrams. In order to reduce computational cost 
for the calculation of four-point $T$ matrix, which describes the dynamical 
electron-hole scatterings resulting in the formation of magnons, the MLWFs are 
utilized. The formalism and its implementation have been extensively detailed in the 
literature \cite{Ersoy-2010,Friedrich2014,Mathias-thesis,Friedrich2019}.

In order to compute smooth altermagnon dispersions along a path through all 
high-symmetry points in the BZ, we employed a $30 \times 30 \times 30$ $\bk$-mesh. 
The response functions were computed at a specific wave vector up to 2.5~eV with 
increments of 5~meV. Goldstone theorem dictates that energy of acoustic magnons 
must approach zero in the long-wavelength limit, or near the $\Gamma$ point of BZ, 
reflecting spontaneous breaking of continuous symmetry in a magnetic system. 
Nevertheless, because of some approximations made in the implementation, this limit 
is usually not satisfied. To fulfill this requirement in our computations, $W$ was 
scaled by a factor of $\lambda \approx 1.24$, i.e., $W \rightarrow \lambda W$.

\begin{figure}[!htb]
\includegraphics[scale=0.7]{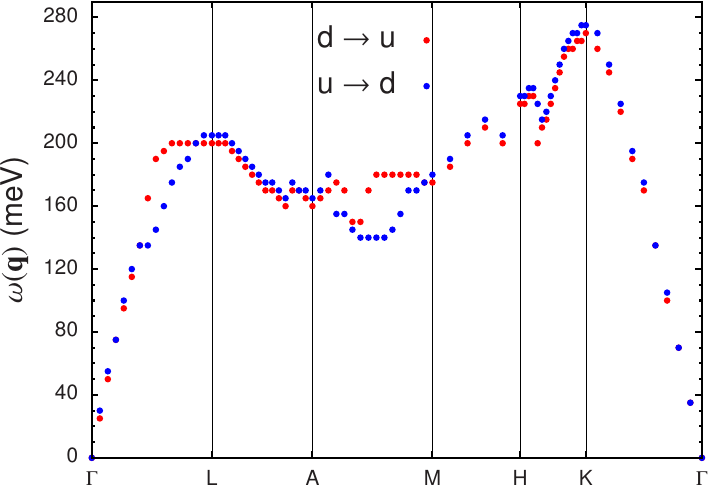}
\caption{The dispersion of altermagnons in bulk CrSb with spin polarizations 
$S_z=\pm 1$.}
\label{pos-du-ud}
\end{figure}

Figure \ref{pos-du-ud} displays unequal energies of altermagnons with opposite spin 
polarizations $S_z=\pm 1$ along the $\Gamma$ - L and A - M directions in the BZ. 
Altermagnons with $S_z=1$ ($S_z=-1$) correspond to excitations of d $\rightarrow$ u 
(u $\rightarrow$ d) spin flips, or spin channels. The energy splitting of 
altermagnons, also called their chirality, directly arises from the spin-split 
electronic bands of AMs. In contrast, conventional AFMs possess degenerate magnon 
modes with identical energies, as observed in other directions. Energy difference 
between these chiral altermagnons reaches 40~meV along the A - M direction. 
Altermagnons display a linear dispersion near the $\Gamma$ point, similar to AFMs. 
They attain an energy of 275~meV (approximately 66.5~THz) at the K point, sharply 
contrasting the GHz-range magnons found in FMs.

\begin{figure}[!htb]
\includegraphics[scale=0.7]{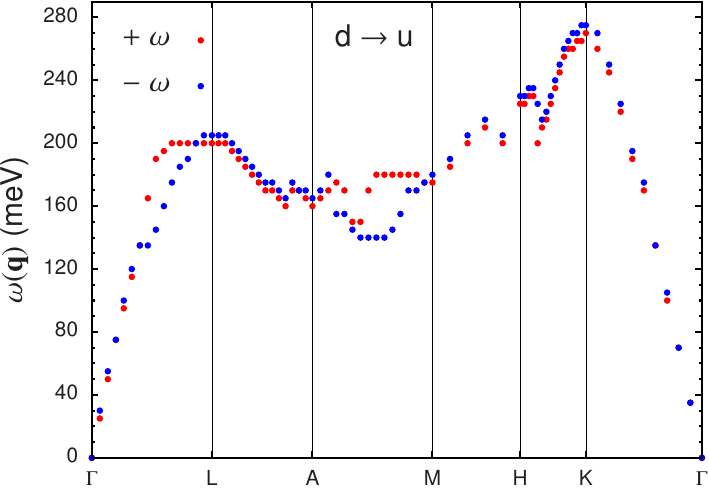} \vskip 0.25cm
\includegraphics[scale=0.7]{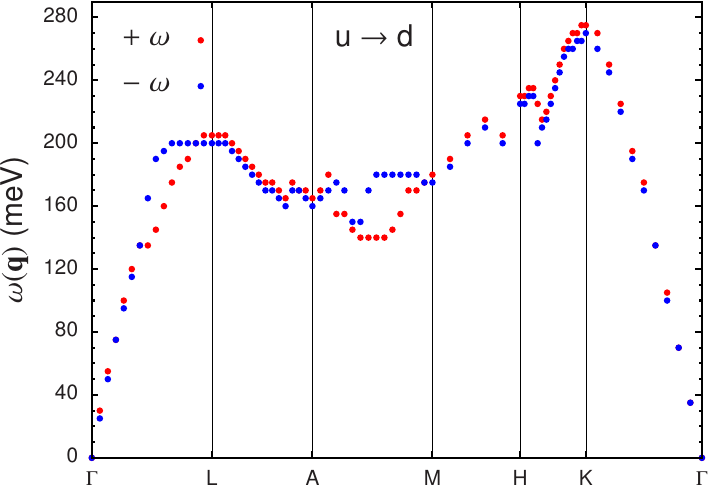}
\caption{The dispersion of altermagnons for d $\rightarrow$ u (top) and 
u $\rightarrow$ d (bottom) spin channels at positive and negative frequencies. 
To facilitate direct comparison, the energies at negative frequencies are 
mirrored onto the positive frequency axis.}
\label{du-ud-pos-neg}
\end{figure}

Subsequently we calculated dispersion of altermagnons at negative frequencies, 
specifically the renormalized Stoner excitations associated with d $\rightarrow$ u 
and u $\rightarrow$ d spin channels. In this case, the response functions were 
calculated at specific $\bq$ points starting from $-2.5$~eV with 5~meV steps. Our 
results are depicted in Fig.~\ref{du-ud-pos-neg}.

In AFMs there is an inherent symmetry $\chi^\mathrm{+-}(-\omega) = 
\chi^\mathrm{-+}(\omega)$, implying that spin excitations at negative frequencies 
correspond to their time-reversed spin counterparts at positive frequencies (e.g., 
an anti-Stoner excitation where a spin-down electron is promoted to a spin-up state). 
Notably, altermagnet bulk CrSb exhibits this symmetry, along with the symmetry 
$\chi^\mathrm{+-}(\omega) = \chi^\mathrm{-+}(-\omega)$ as evident from comparing 
Figs.~\ref{pos-du-ud} and \ref{du-ud-pos-neg}.

\begin{figure}[!htb]
\includegraphics[scale=0.64]{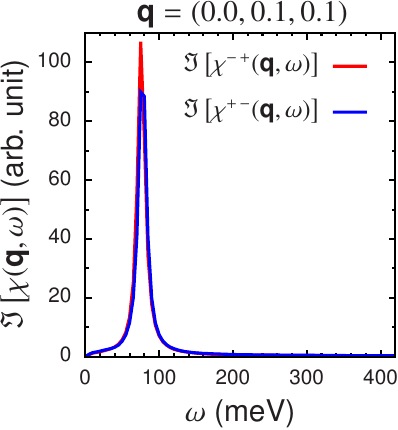}
\includegraphics[scale=0.64]{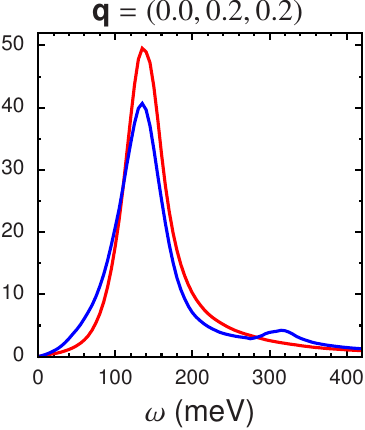} \vskip 0.25cm
\includegraphics[scale=0.64]{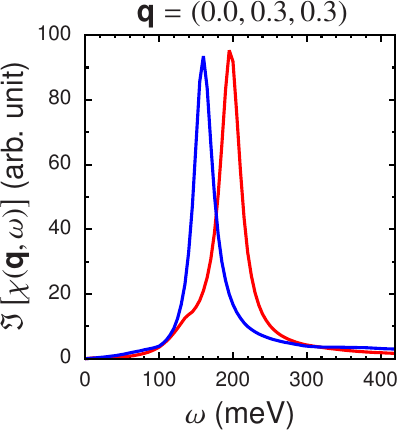}
\includegraphics[scale=0.64]{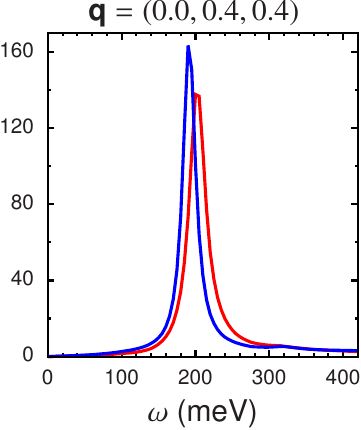}
\caption{The altermagnon spectra along the $\Gamma$ - L direction.}
\label{G-L-TMS}
\end{figure}

\begin{figure}[!htb]
\includegraphics[scale=0.64]{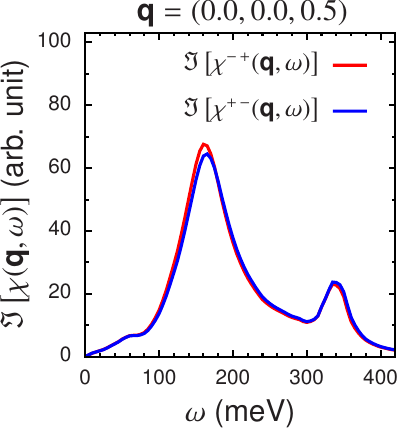}
\includegraphics[scale=0.64]{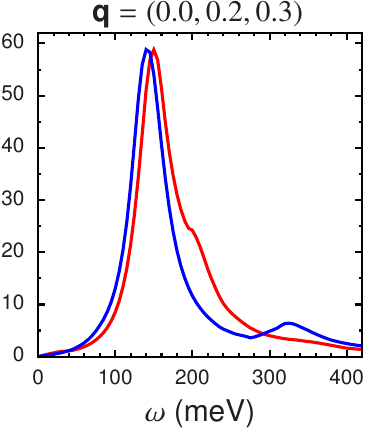} \vskip 0.25cm
\includegraphics[scale=0.64]{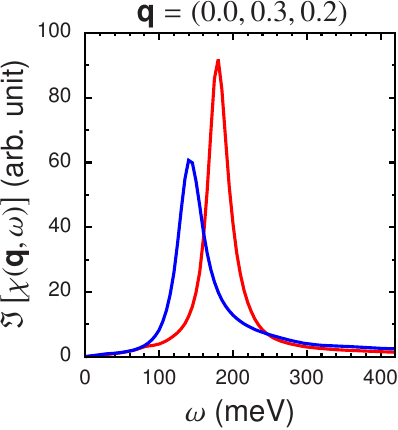}
\includegraphics[scale=0.64]{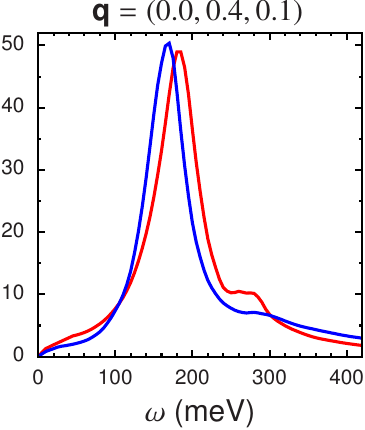}
\caption{The altermagnon spectra along the A - M direction.}
\label{A-M-TMS}
\end{figure}

Magnon dispersion is directly determined by inelastic neutron scattering (INS), 
which maps resonance peaks in the neutron-scattering cross section as a function of 
momentum transfer. The imaginary parts of our computed TMSs for two spin channels 
along the $\Gamma$ - L and A - M directions are respectively plotted in 
Figs.~\ref{G-L-TMS} and \ref{A-M-TMS}. Altermagnon peak characteristics vary with 
wave vector clearly. At small wave vectors, peaks in both spin channels are narrow 
and identical. As $\bq$ increases, however, peak intensities and energies diverge 
between spin channels. Along the A - M direction, satellite peaks appear at higher 
energies as a consequence of strong exchange splitting.


The lifetimes of magnons, or spins, are theoretically estimated through analysis of 
the full width at half maximum (FWHM) amplitude of the imaginary part of the TMS. 
This method relies on the inverse proportionality between the FWHM amplitude of a 
spectral peak and lifetime of corresponding excitation.

As already observed in Figs.~\ref{G-L-TMS} and \ref{A-M-TMS}, altermagnons with 
small wave vectors exhibit narrow peaks and weak damping, resulting in long 
lifetimes. At $\Gamma$ point they are long-living with vanishing energy. However, 
as the wave vector increases, the altermagnon dispersion enters a region where 
density of Stoner excitations is high. This leads to increased decay mechanism and 
shorter lifetimes for altermagnons.

\begin{figure}[!htb]
\includegraphics[scale=0.7]{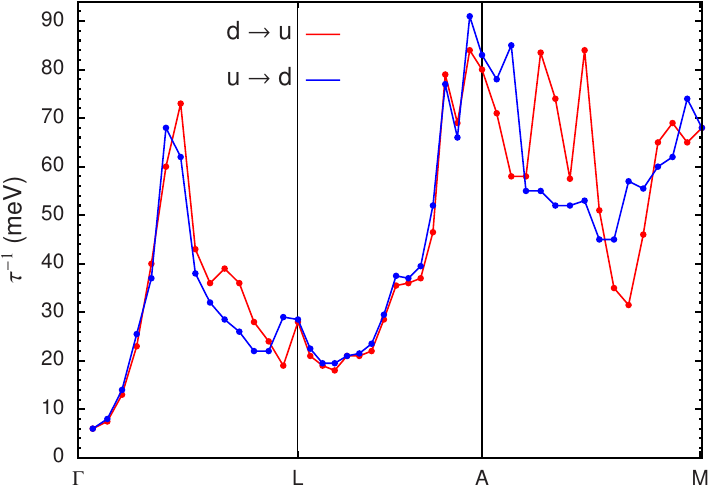}
\caption{The spin relaxation rate of altermagnons in CrSb along the $\Gamma$ - M 
direction, as determined from the FWHM amplitude of imaginary parts of 
$\chi^{-+}(\bq,\omega)$ and $\chi^{+-}(\bq,\omega)$.}
\label{spinlifetime}
\end{figure}

Our analysis, illustrated in Fig.~\ref{spinlifetime}, reveals that lifetimes of 
altermagnons for both d $\rightarrow$ u and u $\rightarrow$ d spin flips exhibit a 
strong dependence on spin channel and propagation direction due to different degrees 
of exchange splitting in the band structure. At a specific $\bq$ point along the 
A - M direction, lifetime differences of altermagnons reach approximately 20~fs, 
corresponding to a difference of 31~meV$^{-1}$ in their inverse decay rates. This 
significant anisotropy is anticipated to have a substantial influence on electronic, 
spin, and energy transport properties, potentially enabling novel applications in 
spintronics.

\section{acknowledgments}
The author is grateful for the kind hospitality of the Peter Gr\"{u}nberg Institut 
(PGI-I) and the Institut for Advanced Simulation (IAS-I) at Forschungszentrum J\"{u}lich 
during this project. M. T. acknowledges the TUBITAK ULAKBIM, High Performance and Grid 
Computing Center (TRUBA resources). He thanks E. {{\c{S}}a{\c{s}}{\i}o{\u{g}}lu and 
C. Friedrich for fruitful discussions.

\section*{Data Availability Statement}
Data available on request from the authors.

\nocite{*}
\bibliography{magnons_in_CrSb.bib}

\end{document}